**IBM Deep Learning Service**

B. Bhattacharjee, S. Boag, C. Doshi, P. Dube, B. Herta, V. Ishakian, K. R. Jayaram, R. Khalaf, A. Krishna, Y. B. Li, V. Muthusamy, R. Puri, Y. Ren, F. Rosenberg, S. Seelam, Y. Wang, J. M. Zhang, L. Zhang

**Abstract**

Deep learning driven by large neural network models is overtaking traditional machine learning methods for understanding unstructured and perceptual data domains such as speech, text, and vision. At the same time, the "as-a-Service"- based business model on the cloud is fundamentally transforming the information technology industry. These two trends: deep learning, and "as-a-service" are colliding to give rise to a new business model for cognitive application delivery: deep learning as a service in the cloud. In this paper, we will discuss the details of the software architecture behind IBM's deep learning as a service (DLaaS). DLaaS provides developers the flexibility to use popular deep learning libraries such as Caffe, Torch and TensorFlow, in the cloud in a scalable and resilient manner with minimal effort. The platform uses a distribution and orchestration layer that facilitates learning from a large amount of data in a reasonable amount of time across compute nodes. A resource provisioning layer enables flexible job management on heterogeneous resources, such as graphics processing units (GPUs) and central processing units (CPUs), in an infrastructure as a service (IaaS) cloud.

# Introduction

The rise of deep learning [1,2] from its roots in neural networks to becoming the state-of-the-art of AI has been fueled by three recent trends: the explosion in the amount of training data; the use of accelerators such as graphics processing units (GPUs); and the advancement in the design of models used for training. These three trends have made the task of training deep layer neural networks with large amounts of data both tractable and useful.

Training deep neural networks, known as deep learning, is currently highly complex and computationally intensive. While GPUs have helped accelerate training, the amount of data as well as complexity of models have increased the computation need beyond the capability of a single GPU. For example, training on 2.5 million images on a single GPU can take 6 days on a simple model [3]. A typical user of deep learning, a data scientist, is also unnecessarily exposed to the details of the underlying hardware and software infrastructure, including configuring expensive GPU machines, installing deep learning libraries, and managing the jobs during execution to handle failures and recovery. Despite the ease of obtaining hardware from infrastructure as a service (IaaS) clouds and paying by the hour, the user still needs to manage those machines, install required libraries, and ensure resiliency of the deep learning training jobs. Furthermore, the user must implement highly complex techniques for scaling and resiliency on their own, as well as keep pace with the updates to the deep learning frameworks in the open source communities.

Instead of being mired with infrastructure and cluster management problems, users would like to focus on training a model in the easiest way possible that satisfies both their cost and performance objectives. This is where the opportunity of deep learning as a service lies. It combines the flexibility, ease-of-use, and economics of a cloud service with the power of deep learning: It is easy to use using the REST APIs, one can train with different amounts of resources per user requirements or budget, it is resilient (handles failures), and it frees users so that they can spend time on deep learning and its applications. Users can choose from a set of supported deep learning frameworks, a neural network model, training data, and cost constraints and then the service takes care of the rest, providing them an interactive, iterative training experience. The job gets scheduled and executed on a pool of heterogeneous infrastructure, including GPUs and CPUs. A simple API (application programming interface) shields users from the complexity of the infrastructure and the advanced mechanics of scaling through distribution. Users can see the progress of their training job and terminate it or modify its parameters based on how it is progressing. When it is done, the trained model is ready to be deployed in the cloud to classify new data.

The value of DLaaS is not limited to data scientists, but extends to developers of new applications and services that would like to add deep learning capabilities but are not able/do not want to build their own software stacks and buy dedicated hardware, or handle scaling and resiliency in-house. Some prominent examples of usage of deep learning within application and services are: speech recognition [4], visual recognition [5], natural language understanding and classification [6], and language translation [7].

IBM DLaaS makes it easy for a provider of such consumer facing cognitive services to provide deep learning training to its users or use it to customize the models in order to provide better outcomes for its customers.

In this paper, we will describe the architecture and experience of the IBM deep learning as a service platform (DLaaS), running in the IBM Cloud. DLaaS was created from the start in close collaboration with deep learning developers across speech, vision, and natural language classification domains. These insights shaped our design, providing guidance into the commonalities between these different workloads and enabled us to provide a cloud service where the infrastructure is shared across these workloads while providing a common API-based access. The rest of the paper describes the user experience of using the DLaaS platform and a study of its usage by about 80 researchers at a workshop, followed by the architecture, and the distribution model. DLaaS is built for extensibility so the paper concludes with a description of how to bring a new framework into the platform. DLaaS is accessible from the IBM Bluemix cloud catalog.

## User Experience and Usage Study

There are four steps that users perform to use DLaaS: (1) prepare their deep learning model, (2) upload the model and training data, (3) start the training job and monitor its progress, and (4) download the trained model once the training job is complete.

The DLaaS interface is designed to be simple and match a user's existing workflow, so step (1) should require minimal effort. The subsequent steps require interacting with the DLaaS Representational State Transfer Application Programming Interfaces (REST) API, either by directly invoking the REST API endpoints, or by using the DLaaS command-line interface (CLI). The CLI provides easy to use command interface over the REST API. The subsequent sub-sections describe the four steps outlined above, with an additional sub-section providing more details about monitoring a running job.

### *Prepare the model*

Users can develop and train their models in their framework of choice (Caffe [8], Torch [9] or TensorFlow [10]) on their local machines, perhaps using a small training dataset. To prepare their models for DLaaS, users must perform three tasks: (1) configure a storage service supported by DLaaS, such as Swift Object Storage, and obtain access credentials; (2) create a manifest.yml file that among other things specifies the deep learning framework they are using and the access tokens to fetch the training dataset from the storage service; and (3) possibly make small changes to their model, such as adjustments to any absolute paths that point to the training dataset. Listing 1 shows an example of a manifest.yml file for a Caffe model.

The manifest can include the resource requirements of the training job, such as the number of learners, and GPUs and memory per learner. These properties can be overridden when a training job is created. The *data_stores* section of the manifest must include a reference to the object storage container that has the training dataset and credentials to access this container, and it may optionally specify the container where the trained model and training logs should be uploaded after the training job as completed. Finally, the framework section points to the main Caffe solver file that defines the model hyperparameters and references the model definition file, as well as any additional arguments to pass to the Caffe.

In the example in Listing 1, a "weights" argument is used to indicate that this model should be incrementally trained by fine tuning the weights in a pre-trained model.

### *Upload the model and data*

Once users have prepared the deep learning model, they must then upload their model to DLaaS and their training dataset to storage service of choice. For the former, the user can either invoke the appropriate DLaaS REST endpoint or use the DLaaS CLI. There are API endpoints to list, create, update, and delete models. The result of deploying the model to DLaaS is a unique generated model ID, which will be used in the next step where a training job is created.

### *Create and monitor a training job*

Once the model has been uploaded, the user can start a training job to train the model. When creating the training job, the user can specify the resource requirements such as the number of learners and the number of GPUs. As with models, there are API endpoints to list, create, and delete training jobs. When a training job is created, a unique training ID is returned and can be used to monitor the progress of the job as detailed in the section below.

### *Download the trained model*
Once the job has completed, the user can download the results, which includes the trained model and a log file that captures the console output during the training job. The results are also optionally stored in the user's storage service, so they can be retrieved directly from there.

### *Understanding Training Progress*
As training jobs can take days or weeks to complete, it is important for a user to monitor its progress to help debug and tune their models and hyperparameters.

Interviews with deep learning users in IBM indicated a number of useful progress indicators: (1) Is the accuracy with expert parameter tuning better than random guessing? (2) Has accuracy hit a plateau, i.e., it is not improving beyond that point? If so, the user would like to be notified and may want to terminate the job. *(3)* Has a state of the model at a certain number of iterations been persisted to checkpoint store? (4) Has a learning rate change after the number of iterations reached a threshold? It is at this point the accuracy jumps, so it may be of interest. (5) Is the accuracy stable for a long time? (6) How often does validation happen and how much time does it take?

The interviews also revealed additional indicators that are relevant to the platform itself, such as detecting idle nodes, and measuring the communication overhead among nodes. These are useful in optimizing the DLaaS platform but are not exposed to the user.

A visualization of deep learning training metrics can be critical in helping unearth insights into the performance of the model and network. A user can notice trends, patterns and anomalies at a glance by visualizing the data, such as understanding when significant improvements or plateaus in the model occurred, or if there are oscillations in the accuracy measure. Such insights, which would be difficult at best to obtain by scrolling through a log file, let the user make quicker and more informed decisions on how to tune their models.

Figure 1 shows an example of ongoing feedback for training progress. This is based on data in logs from the frameworks themselves (e.g. Caffe or Torch). Similar plots with data logs from tools such as nvidia-smi (NVIDIA System Management Interface program) and sysstat/iostat are useful in understanding the training progress and resource utilization. In DLaaS, users can download these logs after the job has completed or stream them to monitor a running job as we discuss later in the paper.

### *DLaaS Usage Study*
We invited around 85 deep learning experts and novices from various universities to use DLaaS in a hands-on colloquium in September 18, 2016 at the T. J. Watson Research Center. Around 75% of the users did not have any prior experience with any deep learning framework. The colloquium was organized in 2 sessions, each with 1.5 hour durations with two different groups of researchers. This provided us with feedback on the usability of service, its scalability and how it can help users to improve accuracy of their models.

During the colloquium, up to 45 users simultaneously started training jobs in DLaaS. Each user submitted at least 1 job and many users submitted 10's of jobs with different resource requirements (e.g., 1, 2, 4 GPUs, different amounts of memory), optimization parameters, etc. DLaaS handled over 200 hundred jobs in a span of three hours. Some jobs finished in a few minutes while others ran for several hours. Outside of the colloquium we also have hundreds of jobs, some that ran over 2 weeks from various researchers.

On the usability aspect, attendees were impressed with how quickly they were able to experiment with deep learning. With infrastructure and cluster management taken care of by the DLaaS system, users were able to begin training a model in minutes, submitting jobs from their own laptops. Many users appreciated that DLaaS was "easy to use", and it was "not complicated to change and redeploy" their jobs. Others liked that DLaaS offers "automatic distribution on multiple nodes", and that the end user can focus on deep learning, and not spend time "having to configure GPUs or handle failure". They liked that they can use custom training data and neural network model defined in one of several supported deep learning frameworks, while DLaaS takes care of hardware and software stack that matches their cost constraints, scalability, and performance requirements.

With respect to the performance aspect, each of users ran using the sample workload (CIFAR10, an established computer-vision dataset used for object recognition) that has a network model with 3 convolutions and 2 fully connected layers. We provided them with hyperparameters that produce about 71% accuracy, and we challenged them to improve the accuracy as much as possible in 1 hour. They ran hundreds of jobs with many parameters changes and fine tuned hyperparameters to achieve an average accuracy of 72.3% while some of them achieved over 77% accuracy. This exercise shows that the DLaaS platform can be used by experts as well as novices to quickly develop and incrementally tune their models until they research the desired accuracy.

We also learned of an interesting gap in our resource and job management. During the colloquium, GPUs of one of the machines became unresponsive but our resource manager failed to recognize this fact and kept scheduling jobs to this node. As a result, a few jobs failed to start because GPUs on that node were not usable. Our resource management layer typically restarts failed jobs but not when the job fails due to either an error in the code or due to a hardware issues like a failed GPU. When the users restarted the failed jobs, they ran successfully.

## DLaaS: Design Principles and Platform Architecture

### *DLaaS Design Principles*

A key design principle of DLaaS is that it provides a large-scale deep-learning platform with multiple GPUs for a learning task by exploiting the economic and technical principles of the cloud paradigm. Contemporary deep learning jobs use high performance computing (HPC) environments with dedicated hardware and failure intolerant, highly customized software stacks. To support deep learning in the cloud, DLaaS addresses the challenges of running on the Cloud such as the dynamic nature of the Cloud, where appropriately handling failures is critical, and exploits the elastic scalability features of Cloud. DLaaS uses a microservices architecture [11] where services are built from the start with resilience in the face of expected failures in the underlying infrastructure. Additionally, cloud services have been supporting mainly short-running or stateless (e.g. Web apps) jobs while deep learning jobs can run for days or even weeks; therefore, DLaaS addresses challenges around data persistence and development and operational issues for non-transactional, data intensive services.

Another key principle of DLaaS is to provide users with the flexibility to use any of the popular deep learning frameworks such as Caffe [8], Torch [9], and TensorFlow [10], combined with the ability to select specific training job service level agreements.

DLaaS also aims to provide a simple and easy to use interface for the users irrespective of the deep learning framework they need to use for the training. To this end, DLaaS provides Application Programming Interfaces (APIs) to prepare the model, to upload the model and training data, to start and get the progress of the training job, and to download the training model. Following these principles, the next section describes the architecture of DLaaS.

## *DLaaS Platform Architecture*

The DLaaS architecture consists of three major components, each deployed as a microservice, as shown in Figure 2. The microservice-based approach [11] enables DLaaS to decompose the core logic into discrete atomic units that can be individually deployed and scaled to handle the traffic.

The *DLaaS API* layer handles all the incoming API requests including load balancing, metering, security and access management. DLaaS *REST API service* instances are dynamically registered into a *service registry* that provides load balancing and fail-over support for incoming API requests. Any jobs that fail are retried automatically a certain number of times before they are marked as failed.

The *DLaaS Core Services* layer is responsible for handling training jobs from submission to completion. This layer consists of five main microservices: (1) A *model deployer service* that handles deploying the model created by the user and persists the model metadata and model input configuration into respective databases. (2) A *Trainer* service that creates a training job out of a given model, (3) A *Lifecycle Manager* (LCM) responsible for deploying training jobs via the trainer service and ensures the progress and resilience of potentially long-running jobs (4) A *Storage Manager* provides reliable connectivity with internal and external storage systems to load this training data and user models from a user-defined store (e.g., a Cloud Storage such as ObjectStore or Network File System (NFS)) and to store the trained models, and (5) A *Metrics Service* collects key metrics that are of interest to the user to understand the training progress and the quality of the training.

The *DLaaS Platform Services* layer provides the key building blocks for executing and managing long running training jobs. At the core is a *GPU-enabled Container Service* that is responsible for executing a training job based on a predefined learner image from the *Docker [12] Registry*. At the time of this writing, we have enhanced the open source Mesos resource manager [13] and Marathon [14] job manager to support GPUs as first class resources. The training jobs request a specified set of resources like number of CPUs, amount of physical memory, number of GPUs and the Mesos/Marathon stack finds the nodes that satisfy these requirements and provisions them for the duration of that job. Container managers such as Kubernetes [15] that can provide the GPU and other resource allocation capability can be used in place of Mesos/Marathon for these jobs. *File Service and Object Store Service* are the primary internal data stores for the training execution to cache the training data, store checkpoints and the trained model. *Apache Zookeeper* [16] (key-value) is used both by DLaaS microservices and by executing training jobs to maintain (minimal) state. Zookeeper is replicated, and is highly available enabling "recovered (after failure) microservice instances" and "training jobs" to continue where their predecessors left off. The DLaaS training jobs as well as the DLaaS platform components are deployed as Docker containers using Docker images available from the *container registry service*. This section describes the architecture of the DLaaS platform – training jobs themselves may rely on other components for synchronization (e.g., the parameter server) which will be described in the next section. A *Logging and Monitoring Service,* which is based on the ELK (Elasticsearch, Logstash and Kibana) stack collects all the logs produced by our services as well as the training jobs and enables users to view the logs of their training jobs for debugging.

In addition to the above components, DLaaS has a real-time visualization component, to enable ease of interaction with long running training jobs (an example shown in Figure 1). This component involves four major aspects: (1) An API that efficiently streams raw logs over a websocket connection. (2) An extensible log parsing API and service, which parses one or more log streams into a common JSON list format. In many cases data points need to be correlated across logs, such as the trainer log, and GPU utilization log from nvidia-smi. Extensibility here allows for the installation of custom parsers to collect and correlate data. (3) The parsed logs are sent to the visualization server, which is currently implemented as a Node.js application using the Express framework. We are exploring a serverless architecture in the near future. (4) The visualization is dynamically rendered on the client in a browser interface using Rickshaw, which is a JavaScript toolkit for creating interactive time series graphs.

# Orchestration of Deep Learning Training Jobs

In this section, we will describe the orchestration mechanisms underlying distributed deep learning training jobs within DLaaS. Requesting a training job results in a call to the *DLaaS Core Services*, which in turn invoke the *DLaaS Platform Services* to fetch the necessary data to start the job, monitor its progress until termination, take actions if the jobs exits or terminates during training and store the trained models back to the user's data store. This section first describes the architecture of a distributed training job (learners, parameter servers and how they interact) before describing how DLaaS orchestrates and manages training jobs: that is, what happens from the time the users request a training job through the DLaaS API until they are able to retrieve their trained model. Figure 3 shows our job orchestration system.

### *Learners*

The key component of distributed deep learning is the use of multiple learner tasks for *data parallelism*. Each learner task is implemented in a (single-machine) deep learning framework (e.g., Caffe, Torch, or TensorFlow) and containerized in a Docker image. Each learner is allotted a configurable number of CPUs and GPUs.

### *Learner Coordination*

DLaaS currently supports the data parallel strategy for workload sharing [17, 18] in contrast to the model parallel strategy or the hybrid strategy [19, 10]. In the data parallelism strategy, each learner has the entire copy of the model (i.e. the entire set of model weights). These model weights are updated locally as a result of training on a new chunk of data. Periodically, learners should synchronize with each other to update their local model by e.g., aggregating model weights from other learners. To perform the model updates while allowing for fault tolerance of the learners, DLaaS uses a parameter server [20] for learners to synchronize periodically and aggregate their weights, as opposed to using broadcast algorithms. Periodically, each learner pushes its weights to the parameter server and pulls updated weights from the parameter server. Weight or parameter aggregation is performed by the parameter server. This leads to a straightforward reduction in the number of messages. In the case of broadcast among all learners (all to all broadcast), the total number of messages exchanged among L learners would be order $L^2$ ($O(L^2)$). With the parameter server, the number of messages exchanged would be order L ($O(L) \approx 2L$), one message from the learner to the parameter with the new model weights and another message from the parameter server to the learner with aggregated weights. Moreover, each learner only needs to be aware of the parameter server (single entity) as opposed to all other learners (L entities), thereby reducing coupling between the learners. The next section provides details about the parameter server used in DLaaS

### *Global Cursor and Work Allocation*

The learners train on data stored in one of the external storage services supported by DLaaS, and collectively make passes over the data set. For each pass, each learner obtains a chunk of data to train on from the external store. Ideally, for each pass over the data set, to ensure data parallelism, each learner operates on a mutually exclusive chunk of data with respect to the other learners. Mutual exclusion is implemented through the use of a global cursor. Each learner computes the size of the data partition that it wants to process, based on its available resources, and assigns itself exclusive data chunks by incrementing the global cursor by the chunk size. Global cursor is implemented through Apache Zookeeper (atomic access and increments to Zookeeper data)

*Parameter Server*
We implemented a parameter server (PS) that is used by the learners to coordinate weights among themselves. As models exhibit a diverse spectrum of training performance under different hardware devices and optimization functions, the parameter server provides several optimization solvers, including parallel stochastic gradient descent (PSGD), elastic averaging SGD, and model averaging, to allow different models to select the most efficient parameter refinement function. Though optimization solvers differ in implementation details, they commonly follow standard iterative convergence algorithms, in which each learner computes local parameters during the forward-pass and back-propagation phases. The forward pass is to assess the quality of existing weights and the back propagation is to generate the gradients with respect to the current weights used by the neural network. Periodically, the learner checks if the condition for parameter push or pull has been satisfied (generally the condition is governed by a communication frequency threshold, for example, a Caffe learner communicates with the PS after 5 batch processing.) The pull function fetches global weights from the server to carry out the next round(s) of iterations. The push function sends locally accumulated gradients or local weights to the parameter server, which then uses a customized aggregation function to update the global weights. Each training job deployed through DLaaS gets its own dedicated parameter server deployment.

The DLaaS parameter server is made up of two key components: (i) a group of parameter server shards that collectively store and aggregate the model parameters from a learning job, and (ii) a PS client library that connects each learner with the parameter server cluster. The PS client is integrated into the learner framework. During training, if the model doesn't fit into the memory (RAM) available on a single machine, the PS client adopts data partitioning to evenly divide the entire model used by the learner based on the number of available servers, and sends partitions to different servers according to the partition ID. As all the learners of the same training job follow exactly the same model partitioning scheme, the same partitions from different learners are gathered by the same server, which then computes a user-specified aggregation function and returns the updated parameters back to the learners. Model partitioning does not imply model-parallelism [19,10]. DLaaS currently only supports data parallelism [17, 18].

As a throughput-critical system, parameter server leverages lockless queues at both network and computation layers to achieve efficient resource utilization. Currently, it relies on TCP/IP protocols to transport data, and uses multithreaded sender/receiver to improve the network throughput. Upon detecting the arrival of a model partition, a receiving thread determines whether it is necessary to invoke the aggregation based on the triggering condition defined by the job. For example, Downpour SGD invokes the aggregation whenever a new partition is received while BSP-based (Bulk Synchronous Parallel programming model) model averaging waits until all partitions are gathered before triggering the aggregation computation. When a job is ready to start the aggregation phase, the aggregation scheduler enqueues the computation into either the CPU-based aggregation queue or the GPU-based queue by taking account of both the estimated aggregation time and potential waiting time of each queue. Eventually, after new parameters are generated, the response threads are invoked to send the new model partition back to all the learners.

The PS client exposes two major synchronous interfaces, called push and pull, to let each learner send out local parameters and retrieve updated models from the server. Additionally, the PS client provides auxiliary functions to manage all the connections (join/leave) to the server. When transferring the data, DLaaS does not use any parameter serialization or deserialization and directly moves all the data in binary format.

## *Lifecycle Management*

As the name suggests, the lifecycle manager (LCM) is the component of DLaaS that is responsible for the entire lifecycle of the training job, from initial deployment to status updates, failure handling and garbage collection of learners and parameter servers. A robust LCM is key to any cloud-based distributed deep learning platform because (i) most public IaaS cloud infrastructures shared between several tenants have a non-trivial rate of failures, network congestions and partitions and because (ii) deep learning jobs are typically long running leaving them susceptible to said failures. The LCM performs the following tasks: (1) Deploys a submitted training job using the Mesos/Marathon cluster management system. (2) Checks whether the parameter server and learners have successfully started. (3) Monitors the status of the learners at runtime and reports their status to the other components. (4) Detects when learners or parameter servers have failed and ensures that failed components are restarted by the cluster management system, and that learning proceeds uninterrupted. (5) Determines when learning has finished so that the training job can be safely terminated and resources allocated to it be reclaimed.

The current implementation of LCM uses the Marathon cluster management system and Apache Zookeeper. The LCM is a micro service that can be independently scaled as needed. It is inherently stateless by itself and stores all state information in Zookeeper. Zookeeper is also used to monitor the status of the learners and parameter server instances. Each container holding either a learner or a parameter server shard is allocated a unique znode (Zookeeper path) by the LCM before deployment; a sidecar (auxiliary) process called the "watchdog" in the container monitors the learner/parameter server and updates its status in the corresponding znode. Status updates can then be read by LCM from Zookeeper. Through status monitoring, the LCM can determine when all learners have finished training, decommission them and reclaim computing resources allocated to them.

Upon job submission, the LCM first deploys the parameter server, and once it has started, queries Marathon to determine the IP address and port on which the parameter server is listening for learners. This information is essential for Learners to connect to the parameter server instances and periodically update learned weights.

### *Single Learner Scenario*

In the case of the training job that contains only one learner, there is no necessity to deploy a parameter server. LCM then deploys the learner, monitors its progress and manages the learner using Zookeeper as described earlier in this section.

### *Fault-Tolerance*
DLaaS is a cloud service so it is expected to be available 24/7, 365 days a year and it is expected to run jobs to completion under scheduled or unscheduled interruptions such as upgrades to the underlying infrastructure, the software stack, failures in various components of the systems as well connectivity issues with the dependent services. Failures in DLaaS can be caused due to faults in DLaaS infrastructure and software stack or due to errors in user input.

Infrastructure faults include physical machine crashes and loss of network connectivity. Faults in the software stack include (i) crashes of containers, (ii) failure of cluster manager (Mesos and Marathon) components and (iii) failure of services on which DLaaS depends on including ObjectStore and Zookeeper. If a node fails, the cluster manager automatically restarts the jobs on that node on a different node. The cluster manager itself is deployed a HA service so unless a majority of the nodes fail, the cluster manager operates without any interruption.

DLaaS microservices, except the storage service, are stateless so they can be upgraded with no impact on running jobs. The storage service handles downloading of the input data and uploading of the resulting models so its upgrades have to coordinated to avoid interruption to the service.

In the case of faults caused by errors in user input, the learner terminates gracefully with appropriate log messages (inside its container), which can be parsed by the "watchdog". The "watchdog", in turn, sends an appropriate status message (JOB_FAILED) to ZooKeeper, which is read by the LCM. The LCM, then updates all pertinent job records in DLaaS with this status, and terminates the job.

All cluster manager components (Mesos and Marathon) checkpoint their state in Zookeeper. If any of these components crash, the recovered containers can continue by using the checkpointed state in Zookeeper. Zookeeper itself is replicated (3-way), and updates to its state are Atomic, strongly Consistent, Isolated and Durable (ACID) due to the use of Zookeeper atomic broadcast. Each learner and parameter server container also creates an ephemeral znode at startup, enabling the LCM to detect learner and parameter server container crashes. By counting the number of active ephemeral znodes, and by reading the status of each active learner from Zookeeper, the LCM can interpret whether the training job is making sufficient progress and whether training can be continued even if a small fraction of learners have failed. After a failure, restarting of the respective containers is handled by Marathon. The LCM also periodically directs learners and parameter servers to checkpoint their state in Object Store. After a failure, recovered learners can start the learning process from a checkpoint, instead of from the beginning. Checkpointing and restart is currently supported for Caffe; we are currently implementing the same for the other frameworks.

The use of Zookeeper decouples the training job (consisting of learners and parameter server instances) from the LCM and other DLaaS microservices, enabling them to fail independently. That is, learning can proceed if any microservice, e.g., the LCM crashed and the LCM can continue accepting training jobs for deployment even if a fraction of currently deployed jobs have failed (e.g., due to an error by the job submitter).

DLaaS microservices are developed to perform exponential backoffs and re-tries for failures associated with network connectivity and access dependent services such as temporary failures in access to Object Store.

As we discussed in the usage study section, our job manager cannot properly restart jobs that gets scheduled to nodes with non-responsive GPU. We are working to periodically check the GPU status and take the node offline in such cases.

**Extensibility**

In a field as dynamic and young as deep learning, there are a lot of diverse deep learning framework used and where the training data is stored. DLaaS provides a pluggable approach to both.

*Integration of other Deep Learning frameworks*

At the time of this writing, we have integrated three well known frameworks: Caffe, Torch, and TensorFlow into DLaaS. DLaaS is designed for pluggability so adding a new DL framework performing vision, speech or other analysis [5, 21] to DLaaS requires nothing more than creating a Docker image that contains three scripts: load.sh, train.sh, and store.sh, which download the training data from a storage service, train the framework-specific model, and upload the trained model to the storage service, respectively.

The framework-specific Docker image should be built on top of a base Docker image we provide that consists of various platform extensions, and a standard set of common libraries for the application like the GPU client libraries and common Operating System (OS) tools. The base image also includes the load.sh and store.sh scripts that interact with different storage systems like a clustered file system, OpenStack Object or block store, or even from other cloud providers.

The custom learner Docker image should include the framework-specific libraries and executables, and an implementation of the train.sh script that will invoke the framework to train a model.

### *Integration of Storage*
An important requirement in deep learning jobs is the management of large training sets. Users may store this data in any number of locations, ranging from local disks, to high performance storage services that support large datasets. DLaaS abstracts access to the external storage service through a pluggable storage component. Currently support for OpenStack Object Storage has been implemented, but others can be easily added. Adding a new storage service to DLaaS involves extending the Storage Manager microservice, updating the load.sh and train.sh scripts in the base learner Docker image, and defining the schema in the model manifest.yml file for the credentials to access the external storage service.

## Related Deep Learning Offerings
Amazon offers Deep Learning Amazon Machine Images (AMI) with several deep learning frameworks that can be launched on Amazon Web Services (AWS) cloud infrastructure. An AMI can be launched on a multi-GPU machine, but the end-user is responsible for setting up CUDA, cuDNN, and other libraries to make use of the GPUs, and the frameworks aren't configured to take advantage of multiple machines. This offering makes it possible to run deep learning workloads on AWS but it not a deep learning service like DLaaS, which abstracts of all the complexity of setup, configuration, etc from the end user. Several other organizations including Microsoft provide pre-configured machines to perform training and relieve some of the setup burden, but these are not cloud-based services. Google offers a more complete distributed deep learning service, but it only supports TensorFlow, not the other frameworks offered by DLaaS.

## Conclusions and Future Work
As deep learning continues to gain traction, the ability to train these complex models easily, economically, and efficiently becomes paramount. This is the goal of DLaaS, the IBM deep learning as a service platform described in this paper that allows a user to train deep learning models in the cloud through REST APIs, without having to leave the comfort of the abstractions and artifacts she is familiar with. We have detailed the architecture of the platform and described the challenges of bringing deep learning to the cloud. We have shown how such a platform can be built in a scalable and resilient manner with data parallelism using a parameter service approach.

There are several open areas to consider from this point forward. We are integrating DLaaS with a machine learning pipeline handling the full lifecycle including data ingestion, data cleaning, inferencing, and so on. In addition, we are making the experience more interactive and enhancing the visualization to deepen the understanding of the training behavior as it progresses in real-time to enable users to react to relevant changes in a timely manner. While supporting end users is important, DLaaS also aims to empower existing cognitive services by allowing them to easily add deep learning capabilities. We are proving this out by integrating with such services. In the future, DLaaS will provide hybrid parallelism and a hyperparameter tuning layer. Such a layer tunes system configuration and training parameters with the goal of improving accuracy while meeting the user's cost and speed needs. Interestingly, DLaaS, as a cloud-based deep learning service, affords the opportunity to learn from the performance and characteristics of previously observed models and training parameters to optimize and offer suggestions to future users.

```
name: my-mnist-model
version: "1.0"
description: Caffe MNIST (Mixed National Institute of Standards and
Technology database) model running on GPUs. The MNIST database (Mixed
National Institute of Standards and Technology database) is a large database
of handwritten digits that is commonly used for training various image
processing systems.
Learners: 2
gpus: 2
memory: 8000MiB

data_stores:
  - id: softlayer-object-storage
    type: softlayer_objectstore
    training_data:
      container: my_training_data
    training_results:
      container: my_training_results
    connection:
      auth_url: https://dal05.objectstorage.softlayer.net/auth/v1.0
      user_name: my-user-name
      password: my-password

framework:
  name: caffe
  version: "1"
  job: lenet_solver.prototxt
  arguments:
    weights: pretrained.caffemodel
```

Listing 1: An example of a manifest.yml file for a Caffe model. The resource requirements such as number of learners, number of GPUs, amount of memory can be overridden when creating a training job.

Figures:

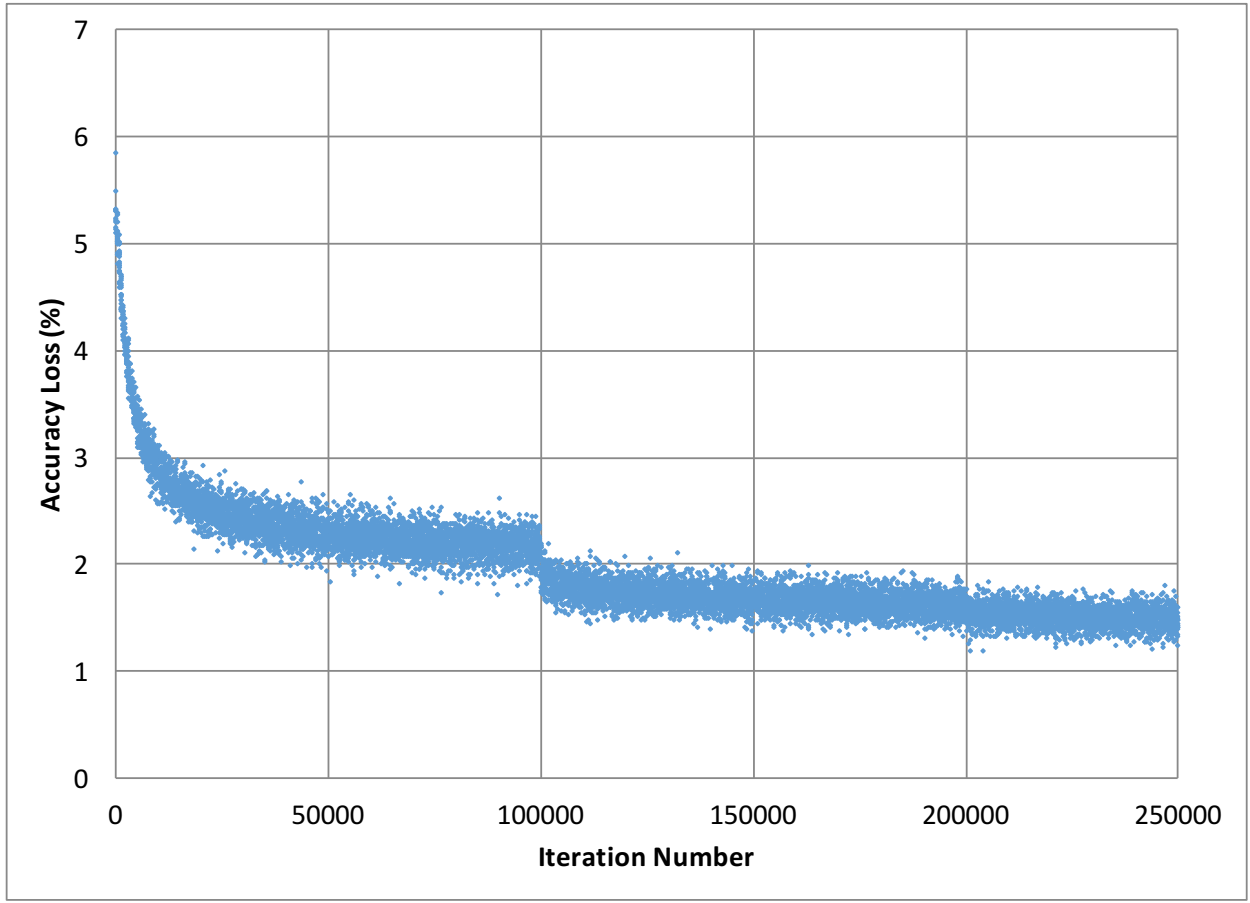

Figure 1: Example visualization provided in DLaaS service. This shows the accuracy loss as the training progress in time. Developers can use this information to decided when to stop the job or at what points to tune the model. In the example they may decide to understand why there is a sudden drop in accuracy loss around 100000 iterations.

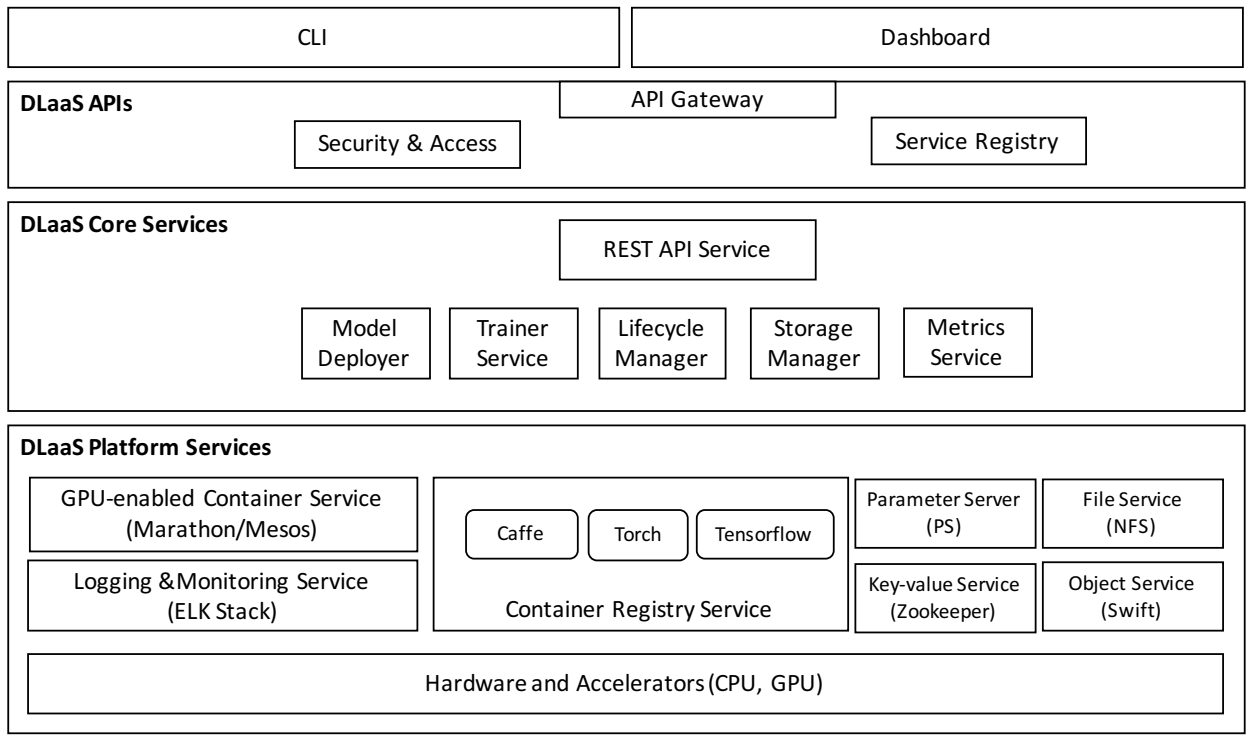

Figure 2 - Overview of the DLaaS Architecture. Users interact with command line interface (CLI) or Dashboard and the core system consists of APIs, a set of microservices for model and training job life cycle management and a set of platform services that provide GPU enabled infrastructure, data services to transfer the training data into the system and to transfer the trained models and logs out of the system, and logging and monitoring services to get the details of the training progress.

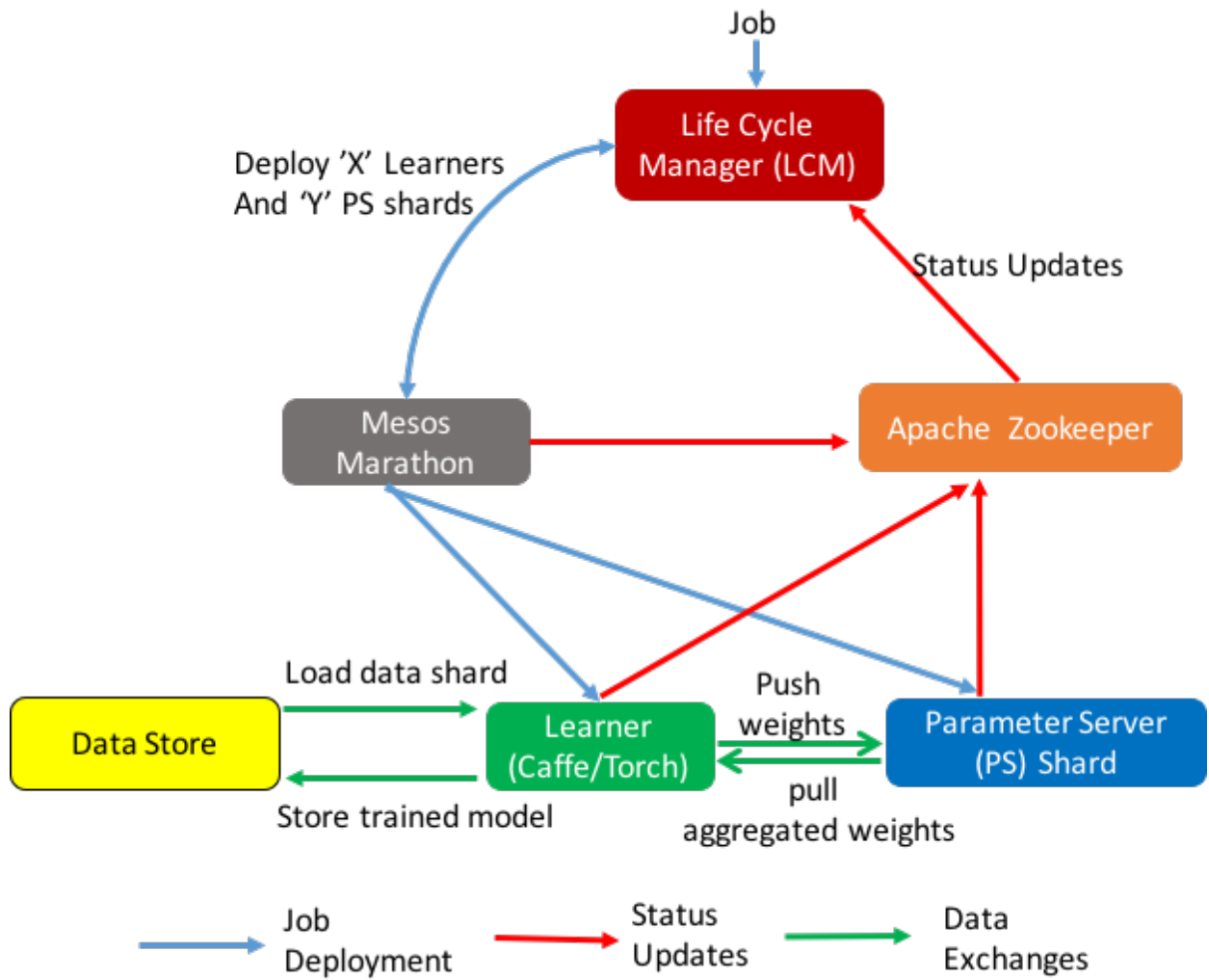

Figure 3: DLaaS Distribution Model. The life cycle manager works with the cloud job manager such as Mesos/Marathon to deploy a collection of parameter servers and learners of a multi-node distributed job and it orchestrates the life cycle via the Zookeeper key value store.

Author Bios

**Bishwaranjan Bhattacharjee** *IBM Research Division, Thomas J. Watson Research Center, Yorktown Heights, New York 10598 (bhatta@us.ibm.com).* Mr. Bhattacharjee is a Senior Technical Staff Member and Master Inventor at the Thomas J. Watson Research Center. His current research interests include scalable data management and deep learning.

**Scott Boag** *IBM Research Division, Thomas J. Watson Research Center, Cambridge MA 02142 (scott_boag@us.ibm.com),* Mr. Boag is a Senior Technical Staff Member at IBM Research. His research interests are in compilers and data transformation.

**Chandani Doshi** *IBM Research Division, Thomas J. Watson Research Center, Cambridge MA 02142 (cdoshi@mit.edu)*. Ms. Doshi is a Computer Science and Electrical Engineering student at the Massachusetts Institute of Technology. She worked on DLaaS during a summer internship at IBM Research.

**Parijat Dube** *IBM Research Division, Thomas J. Watson Research Center, Yorktown Heights, NY 10598 (pdube@us.ibm.com)*. Dr. Dube is a Research Staff Member in Cloud and Cognitive Platform department at the IBM T. J. Watson Research Center. He received his PhD (2002) in computer science from INRIA, France. His research interests are in performance modeling, analysis, and optimization of systems.

**Ben Herta** *IBM Research Division, Thomas J. Watson Research Center, Yorktown Heights, New York 10598 (bherta@us.ibm.com)*. Mr. Herta is a Senior Software Engineer interested in ensuring programs run efficiently on large-scale systems, taking advantage of specialized hardware such as Infiniband and GPUs especially for cognitive workloads.

**Vatche Ishakian** *IBM Research Division, Thomas J. Watson Research Center, Cambridge MA 02142 (vishaki@us.ibm.com)*. Dr. Ishakian is a Research Staff Member and a Visiting Fellow at Boston University. His interests are in cloud computing, resource management, application-level scheduling, network optimization and economics, data placement, and network architecture.

**K. R. Jayaram** *IBM Research Division, Thomas J. Watson Research Center, Yorktown Heights, New York 10598 (jayaramkr@us.ibm.com)*, Jayaram is a Research Staff Member at IBM Research. His research interests include distributed systems, cloud infrastructure, and distributed data analytics platforms.

**Rania Khalaf** *IBM Research Division, Thomas J. Watson Research Center, Cambridge MA 02142 (rkhalaf@us.ibm.com)*. Dr. Khalaf is a Distinguished Research Staff Member and Senior Manager at IBM Research. Her research is at the intersection of Cloud Computing, Service Composition and Machine/Deep Learning.

**Avesh Krishna** *IBM Research Division, Thomas J. Watson Research Center, Yorktown Heights, New York 10598*. (ajk11@rice.edu) Mr. Krishna is a Computer Science and Political Science student at Rice University. He worked on DLaaS during an internship at IBM Research.

**Yu Bo Li** *IBM Research Division, China Research Laboratory, Haidian District, Beijing, China 100193 (liyubobj@cn.ibm.com)*. Dr. Li is a Research Staff Member at IBM Research - China. His current research interests include GPU enablement and optimization on cloud, deep learning framework, and container technology.

**Vinod Muthusamy** *IBM Research Division, Thomas J. Watson Research Center, Yorktown Heights, New York 10598 (vmuthus@us.ibm.com)*. Mr. Muthusamy is a Research Staff Member at the Thomas J. Watson Research Center. His current research interests include cloud platforms that support a variety of workloads, programming models and technologies to compose services, and analytic tools to monitor and debug distributed applications.

**Ruchir Puri** *IBM Watson, Yorktown Heights, New York 10598 (ruchir@us.ibm.com)*. Dr. Puri is an IBM Fellow and Chief Architect of IBM Watson where he is responsible for architecture across the range of Watson offerings. He led Deep Learning and Machine Learning Platform Initiative at IBM Research and drove IBM's strategy for differentiated cognitive computing infrastructure.

**Yufei Ren** *IBM Research Division, Thomas J. Watson Research Center, Yorktown Heights, New York 10598 (yren@us.ibm.com)*. Dr. Ren is a Research Staff Member at IBM T. J. Watson Research Center, working on system performance optimization and large-scale machine learning system development. Ren's research interests span the areas of high performance networks, storage systems, I/O performance optimization, and parallel and distributed computing.

**Florian Rosenberg** *IBM Austria, Obere Donaustrasse 95, 1020 Vienna, Austria (rosenberg@at.ibm.com)*. Dr. Rosenberg is a Senior Technical Staff Member at IBM Austria. His current interests include cloud architectures and platforms, programming and deployment models to deliver high-quality cloud services across different industry segments.

**Seetharami R. Seelam** *IBM Research Division, Thomas J. Watson Research Center, Yorktown Heights, New York 10598 (sseelam@us.ibm.com)*. Dr. Seelam is a Research Staff Member at the T. J. Watson Research Center. His current research interests include developing technology to deliver hardware, middleware, containers, and applications as-a-service on the cloud.

**Yandong Wang** *IBM T. J. Watson Research Center, Yorktown Heights, NY 10598 USA (yandong@us.ibm.cim)*. Dr. Wang is a Research Staff Member in the Cognitive System Analysis and Optimization Group. His current interests include building large-scale computing platform for big data analytics and machine learning algorithms.

**Jian Ming Zhang** *IBM Research Division, China Research Laboratory, Haidian District, Beijing, China 100193 (zhangjm@cn.ibm.com)*. Mr. Zhang is a Research Staff Member in the Cloud Services department at IBM Research - China. His current research interests include cloud platform and DevOps, service management, and operational analytics.

**Li Zhang** *IBM T. J. Watson Research Center, Yorktown Heights, NY 10598 USA (zhangli@us.ibm.com)*. Dr. Zhang is a Master inventor, Principal Research Staff Member, and Manager of the Cognitive System Analysis and Optimization Group at the IBM T. J. Watson Research Center. His current interests include large-scale Big Data and Machine learning systems.